\documentclass[12pt,usenames,dvipsnames,a4paper]{article}

\usepackage[utf8]{inputenc}
\usepackage[T1]{fontenc}

\usepackage{changepage}
\usepackage{amsmath,amsfonts,amssymb}
\usepackage{epsf,amsmath,bbold,amsfonts,stmaryrd}
\usepackage{mathrsfs}
\usepackage{appendix}
\usepackage{caption}
\usepackage{color}
\usepackage{datetime}
\usepackage{float}
\usepackage{graphicx}
\usepackage[colorlinks,hyperindex,breaklinks]{hyperref}
\hypersetup{pageanchor=false,citecolor=red,urlcolor=red}
\usepackage{indentfirst}
\usepackage[numbers,square,comma,sort&compress,merge]{natbib} 
\usepackage{subfig}
\usepackage{mathtools}
\usepackage{ytableau}
\usepackage{tabu}
\usepackage{esvect}

\usepackage{calc}
\usepackage{bm}

\allowdisplaybreaks

\def\a{\alpha}

\def\b{\beta}
\def\c{\chi}

\def\d{\delta}

\def\f{\frac}

\def\l{\left}

\def\mc{\mathcal}

\def\m{\mu}
\def\n{\nu}

\def\p{\partial}

\def\r{\right}
\def\s{\sigma}

\def\x{\xi}

\def\be{\begin{equation}}
\def\ee{\end{equation}}

\def\bea{\begin{eqnarray}}
\def\eea{\end{eqnarray}}

\def\ba{\begin{array}}
\def\ea{\end{array}}

\def\bc{\begin{center}}
\def\ec{\end{center}}

\def\bl{\begin{flushleft}}
\def\el{\end{flushleft}}

\def\br{\begin{flushright}}
\def\er{\end{flushright}}

\def\bi{\begin{itemize}}
\def\ei{\end{itemize}}

\def\bt{\begin{tabular}}
\def\et{\end{tabular}}

\makeatletter

\newsavebox\myboxA
\newsavebox\myboxB
\newlength\mylenA

\newcommand*\xoverline[2][0.75]{%
    \sbox{\myboxA}{$\m@th#2$}%
    \setbox\myboxB\null
    \ht\myboxB=\ht\myboxA%
    \dp\myboxB=\dp\myboxA%
    \wd\myboxB=#1\wd\myboxA
    \sbox\myboxB{$\m@th\overline{\copy\myboxB}$}
    \setlength\mylenA{\the\wd\myboxA}
    \addtolength\mylenA{-\the\wd\myboxB}%
    \ifdim\wd\myboxB<\wd\myboxA%
       \rlap{\hskip 0.5\mylenA\usebox\myboxB}{\usebox\myboxA}%
    \else
        \hskip -0.5\mylenA\rlap{\usebox\myboxA}
         {\hskip 0.5\mylenA\usebox\myboxB}%
    \fi}
\makeatother

\def\be{\begin{equation}}
\def\ee{\end{equation}}

\def\bea{\begin{eqnarray}}
\def\eea{\end{eqnarray}}

\def\f{\frac}

\def\p{\partial}

\usepackage{xspace}

\newcommand*{\eg}{e.g., }

\usepackage{xifthen}
\newcommand*\diff{\mathrm{d}} 
\newcommand*\ldiff[2][]{ \ifthenelse{\isempty{#1}}{ \diff
#2}{\diff^#1#2} \,} 
\let\limitint\int 
\renewcommand{\int}{\limitint \!} 

\begin{document}

\begin{titlepage}

\vspace*{-1cm}

\begin{center}
\Large\textbf{QCD axion from broken scale symmetry}
\end{center}

\begin{center}
\textsc{Georgios K. Karananas\,$^{\star,\dagger}$~Mikhail
Shaposhnikov\,$^\ddagger$}
\end{center}

\begin{center}
\it {$^\star$Max-Planck-Institut für Physik\\
Boltzmannstraße 8, 85748 Garching bei M\"unchen, Germany\\
\vspace{.4cm}
$^\dagger$Arnold Sommerfeld Center\\
Ludwig-Maximilians-Universit\"at M\"unchen\\
Theresienstra{\ss}e 37, 80333 M\"unchen, Germany\\
\vspace{.4cm}
$^\ddagger$Institute of Physics \\
\'Ecole Polytechnique F\'ed\'erale de Lausanne (EPFL) \\ 
CH-1015 Lausanne, Switzerland\\
\vspace{.4cm}
}
\end{center}

\begin{center}
\small
\texttt{\small georgios.karananas@physik.uni-muenchen.de}  \\
\texttt{\small mikhail.shaposhnikov@epfl.ch} 
\end{center}

\vspace*{2cm}

\begin{abstract} 

A consistent non-compact axion cosmology requires a non-periodic field, an
effective field theory valid sufficiently above the inflationary scale, and a
small non-QCD contribution to the potential that tilts the axionic vacuum
landscape in order to trigger a timely domain-wall collapse. All conditions can
be met by the dilaton---the pseudo-Nambu-Goldstone boson of spontaneously
broken approximate scale invariance. 

\end{abstract}

\end{titlepage}

\section{Introduction}

Stripped to its bare essentials, the axion
solution~\cite{Peccei:1977hh,Weinberg:1977ma,Wilczek:1977pj} to the strong CP
puzzle only needs a light dynamical scalar degree of freedom---not necessarily
compact~\cite{Karananas:2025uhy}---coupled to the QCD topological density. 

A concrete example of a non-compact axion was presented
in~\cite{Karananas:2025uhy}, motivated by Weyl-invariant Einstein-Cartan
gravity~\cite{Karananas:2024xja}. There, the resulting cosmological dynamics
was shown to be qualitatively distinctive. Inflationary fluctuations of the
effectively massless field populate many neighboring branches of the eventual
QCD potential. When QCD turns on, the axion density on Cosmic Microwave
Background (CMB) scales is averaged over many uncorrelated vacua, thereby
evading the usual isocurvature bound. Since no global U(1) symmetry is
invoked, there are no cosmic strings, so the defect sector consists only of
domain walls. Cosmological viability requires a small non-QCD bias that lifts
the degeneracy among adjacent ground states and triggers the collapse of the
wall network. As shown in~\cite{Karananas:2025uhy}, this implies two
correlated signatures: non-vanishing residual strong CP violation and a
stochastic gravitational-wave background peaking at nanohertz frequencies.

The purpose of the present work is to show that spontaneously broken exact
or approximate scale invariance provides yet another microscopic origin for
precisely this kind of effective dynamics. The reason is conceptually simple,
and the picture that emerges is rather economical. Non-linearly realized
dilatations already furnish a non-compact scalar degree of freedom, the
dilaton. At the same time, sources of explicit scale-symmetry breaking
generically induce a non-QCD contribution to the potential of the dilaton.
Thus, two of the structural ingredients required by the non-compact axion
scenario emerge together. 

For gravity to enter the picture---its presence is absolutely essential for
our considerations here---scale invariance requires the usual nonminimal
coupling of the dilaton to the Ricci scalar. Once one passes to the Einstein
frame the dynamics becomes especially transparent, with the canonical axion
field related to the original compensator via a logarithmic mapping, and thus
scale invariance being nonlinearly realized as an exact or approximate shift
symmetry.

It is important to point out that the microscopic source of breaking matters,
and different realizations can lead to different deformations. This ambiguity
is resolved in a particularly compelling way when dilatations are promoted to
a quantum symmetry and gravity is formulated as a
unimodular~\cite{Shaposhnikov:2008xb} or transverse-diffeomorphism (TDiff)
theory~\cite{Blas:2011ac,Karananas:2016grc}. In that setting, no explicit
scale-breaking operators are radiatively generated when the theory is
renormalized in a scale-invariant
manner~\cite{Englert:1976ep,Shaposhnikov:2008xi}. The sole non-QCD
contribution to the dilaton potential is then a \emph{uniquely fixed}
exponential in the Einstein
frame~\cite{Shaposhnikov:2008xb,Garcia-Bellido:2011kqb}. As we show, the same
potential appears for a broad class of explicit scale-breaking operators; see
also~\cite{Csaki:2014bua} for related considerations in inflation.

Equally important is the communication of the dilaton to QCD. A logarithmic
coupling of the field to the topological density is perfectly compatible with
scale invariance, since under dilatations this shifts only by a total
derivative. Interestingly, in a scale-invariant renormalization prescription,
where the regulator is replaced by (a function of) the dilaton, the
renormalized strong CP parameter in the presence of weak CP
violation~\cite{Ellis:1978hq} induces precisely this type of operator. From
the low-energy point of view, its coefficient---which controls the axionic
decay constant---is simply a free parameter.

This paper is organized as follows. In Sec.~\ref{sec:setting}, we set up the
theory. In Sec.~\ref{sec:nonlinear}, we discuss its broken phase. In
Sec.~\ref{sec:pheno}, we outline the main phenomenological consequences. We
conclude in Sec.~\ref{sec:conclusion}.

\section{Setting the stage}
\label{sec:setting}

The two-derivative Lagrangian comprising a dilaton $\c$ coupled to gravity and
QCD, that is classically invariant under linearly realized global dilatations
\be
\label{eq:dilatations_linear}
g_{\m\n} \mapsto e^{-2\s}g_{\m\n} \ ,
~~~\c\mapsto e^\s\c \ ,
~~~\s = {\rm constant} \ ,
\ee
reads
\be
\label{eq:Lagrangian_dilaton_exact}
\mc L = \f{\x_\c \c^2}{2} R - \f 1 2 (\p_\m \c)^2 
- \a \c^4 + \beta \log \left(\f{\c}{M}\right) Q \ .
\ee

Even though the dilaton was explicitly introduced in the above, note that it
may have a purely gravitational origin. In scale-invariant TDiff
theories~\cite{Blas:2011ac}, for instance,  the determinant of the metric is a
dynamical degree of freedom playing the role of $\chi$. Yet another example is
pure $R^2$ gravity, which is classically scale invariant and propagates,
besides the massless graviton, an extra scalar mode, see
\eg~\cite{Alvarez-Gaume:2015rwa}.

For brevity, in~(\ref{eq:Lagrangian_dilaton_exact}) we kept only the terms
relevant for the present discussion, suppressing the standard kinetic term for
glue, as well as Standard Model (SM) fields and their possible interactions
with the dilaton; apart from a nonderivative coupling of $\c$ to the Higgs
field, such terms generally involve $\p_\m\c/\c$ to appropriate powers.
$\x_\c$, $\a$ and $\beta$ are dimensionless, while $M$ is an arbitrary mass
scale. The QCD topological density is
\be
\label{eq:QCD_topological}
Q = \f{\a_s}{8\pi}G_{\m\n}^b \widetilde G^{b\,\m\n} \ ,
~~~\a_s = \f{g_s^2}{4\pi} \ ,
\ee
with $G_{\m\n}$ the SU(3) field-strength tensor, $\widetilde G_{\m\n}$ its
dual, $g_s$ the corresponding strong coupling, and summation over the color
index $b$ is assumed. Notice that under~(\ref{eq:dilatations_linear}), the
logarithmic coupling of the dilaton to~(\ref{eq:QCD_topological}) responds as  
\be
\label{eq:dilaton_Q_coupling}
\log \left(\f{\c}{M}\right) Q \mapsto 
\log \left(\f{\c}{M}\right) Q + \s Q \ ,
\ee
meaning that the variation is proportional to $Q$ itself. Since $Q$ is related
to the SU(3) Chern-Simons current $K_\m$ as $Q= \nabla_\m K^\m$, the
Lagrangian classically shifts by a total derivative, so this term is perfectly
admissible despite the appearance of the explicit mass scale $M$. This can
also be seen from the equations of motion following
from~(\ref{eq:Lagrangian_dilaton_exact}), which are independent of $M$.
Equivalently, the dilaton-QCD interaction can be written as 
\be
\label{eq:shift_symmetric_Kmu}
-\beta \frac{\partial_\mu\chi}{\chi} K^\mu \ , 
\ee 
which is obviously invariant under the
transformation~(\ref{eq:dilatations_linear}). Although $K^\mu$ itself is not
gauge invariant, the action is invariant under ``small'' gauge
transformations with zero topological charge---the only requirement needed for
the consistent quantization of the theory. 

Let us also comment on massive fermions. At the fundamental level quark masses
arise from scale-invariant Yukawa interactions with the SM Higgs. Their
complex phases feed into the physical strong CP angle, but they do not
introduce an additional $\c$-dependent contribution to $\bar\theta$. Of
course, derivative couplings of the dilaton to fermionic currents are allowed.
If such terms are integrated by parts and the anomalous Ward identity is used,
one obtains both a contribution proportional to $G\widetilde G$ and the
fermion mass. These terms, however, better be kept together, as they are
simply a different operator basis for the original shift-invariant derivative
interaction~\cite{Karananas:2025ews}.

A further very important point concerns the logarithmic coupling of the
dilaton to $Q$. This is obviously admissible by symmetry arguments--- there
are no compelling reasons why scale-invariant CP-odd operators should not be
included. After all, CP is a discrete symmetry already broken in the SM.
Moreover, its presence is in fact unavoidable. It was
shown~\cite{Ellis:1978hq} long ago that in the presence of weak CP violation,
the theta parameter gets renormalized, and later
analysis~\cite{Khriplovich:1993pf} clarified the corresponding ultraviolet
divergence and its relation to the physical CP-odd invariant, the Jarlskog
determinant~\cite{Jarlskog:1985ht}, of the theory. If one adopts a
scale-invariant regularization
prescription~\cite{Englert:1976ep,Shaposhnikov:2008xi}, in which the
subtraction scale is replaced by (a function of) the dilaton, the CP-odd
counterterms that renormalize the strong phase acquire an evanescent
dependence on $\chi$, which necessarily induces contributions to the effective
action of the form $\log \c \,Q$. This can be understood in a simple way.
Work, for instance, in dimensional regularization. Schematically, a divergent
contribution which in an ordinary scheme would renormalize the coefficient of
$Q$ is now replaced by
\be
\f{1}{D-4} \chi^{D-4} Q \ .
\ee
Expanding this around $D=4$ spacetime dimensions, gives
\be
\f{1}{D-4}Q +\log\left(\f{\chi}{M}\right)Q +\ldots \ ,
\ee
with $M$ an arbitrary and unphysical scale introduced only to make the
argument dimensionless; the ellipses stand for terms proportional to $D-4$.
After subtracting the pole, the operator that survives at $D\to 4$ is a finite
logarithmic dilaton-$Q$ interaction. Its strength---controlled by the
parameter $\b$---is of course not computable within the effective theory.

\begin{figure}[!t]
    \centering
    \includegraphics[scale=.5]{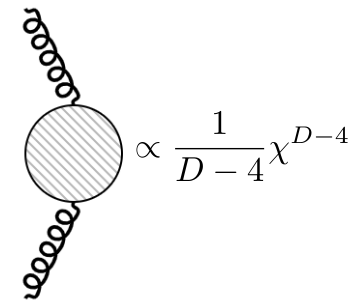}
    \caption{Diagrammatic origin of the logarithmic dilaton-QCD coupling. Weak
CP-violating quark/Yukawa dynamics, represented here by a blob, induces a
divergent CP-odd counterterm with two external gluons. In the scale-invariant
regularization prescription, this counterterm is dressed by an evanescent
compensator factor, $\chi^{D-4}$. Expanding around $D=4$ and subtracting the
pole $\propto 1/(D-4)$ leaves a finite local operator proportional to
$\log\c\,Q$ in the limit $D\to 4$.}
    \label{fig:Feynman}
\end{figure}

One may also think of this diagrammatically, see Fig.~\ref{fig:Feynman}, as
the usual weak-CP-induced renormalization of the QCD vacuum angle, with
external gluons and CP-violating quark/ Yukawa insertions. The scale-invariant
regularization, however, converts any evanescent dependence on the regulator
into a finite logarithmic coupling. A closely related perturbative generation
from evanescent operators was discussed in~\cite{Karananas:2025ews}.

When nonperturbative QCD dynamics kicks in, the vacuum is endowed with a
non-vanishing topological susceptibility, reflected in the fact that the
two-point function of $K_\m$ has a pole at vanishing
virtuality~\cite{Luscher:1978rn}. Consequently, the would-be surface term is
saturated in the infrared and the constant shift~(\ref{eq:dilaton_Q_coupling})
becomes physical, as the vacuum energy depends nontrivially on the coefficient
that multiplies $Q$. At first sight, this may seem to be in tension with scale
invariance. In spite of what one may naively conclude, dilatations are
actually realized by the QCD vacuum in a subtle way\,\footnote{For a
physically transparent discussion with details
see~\cite{Dvali:2005an,Dvali:2017mpy,Dvali:2022fdv}.}: QCD induces the
familiar periodic potential for the quantity that multiplies $Q$, with
constant shifts compensated by a corresponding relabeling of the vacuum
branch~\cite{Dvali:2005an}.~The symmetry is only broken once a vacuum is
selected, see~\cite{Karananas:2026stg} for related discussions. Equivalently,
the vacuum energy depends only on the invariant combination of the
$\bar\theta$ parameter with the field multiplying $Q$. Equivalently, if the
action is written as in~(\ref{eq:shift_symmetric_Kmu}), the scale symmetry is
manifest. Despite the fact that it is spontaneously broken, the Goldstone
theorem---telling that there is a massless Goldstone boson---does not apply,
since the dilaton is coupled to the current $K^\mu$, the correlator of which
contains a massless pole. This leads to the mass generation of the physical
excitation associated with $\chi$, which is nothing but the scale-invariant
combination $\log{\chi(x)}-\log{\chi(@\rm{boundary})}$. In this way, the
dilaton can play the role of the axion and relax strong CP violation to
experimentally acceptable levels.

So far so good, but the question of cosmological viability still remains. The
QCD-induced potential comprises a degenerate set of vacua, so unless this
degeneracy is somehow lifted there is no preferred ground state and the
corresponding domain-wall network is stable. The classically scale-invariant
quartic self-interaction $\propto\c^4$ is of no use here, as it is a
cosmological constant in disguise; this will become clear shortly once we move
to the ``Einstein'' frame, where gravity is canonical and the symmetry is
nonlinearly realized.

The fate of classical scale invariance at the quantum level can be different.
Depending on how radiative corrections are treated, it may be broken by matter
loop effects, or instead be required to remain
intact~\cite{Englert:1976ep,Shaposhnikov:2008xi}. Even in the latter case,
however, additional non-derivative pieces in the dilaton potential are
generically expected. This is the state of affairs in unimodular
gravity~\cite{Shaposhnikov:2008xb,Garcia-Bellido:2011kqb},
transverse-diffeomorphism theories~\cite{Blas:2011ac}, and three-form
embeddings~\cite{Karananas:2026stg}. 

In addition, beyond two derivatives, the effective action for the dilaton
contains a number of terms (see
\eg~\cite{Shaposhnikov:2022dou,Shaposhnikov:2022zhj}), among which two stand
out, as they are logarithmic couplings of $\c$ to the gravitational
topological densities
\be
\label{eq:Jordan_topological}
\mc L_{\rm grav} \supset 
\eta_E \log\l(\f{\c}{M}\r) E_4 + \eta_P \log\l(\f{\c}{M}\r) P_4 \ ,
\ee
where the $\eta$'s are dimensionless; $E_4$ and $P_4$ are the
Euler/Gauss-Bonnet and Pontryagin invariants, respectively: 
\be
\label{eq:gravity_topological}
E_4 = \f 1 4\epsilon_{\a\beta\gamma\delta} \epsilon^{\m\n\rho\s}
R^{\a\beta}_{~~~\m\n} R^{\gamma\delta}_{~~~\rho\s} \ , 
~~~P_4 = \f 1 2 \epsilon_{\a\beta\gamma\delta} 
R^{\a\beta}_{~~~\m\n} R^{\gamma\delta\m\n} \ .
\ee 
Since $E_4$ and $P_4$ themselves are total derivatives, the variation
of~(\ref{eq:Jordan_topological}) under the constant
dilatation~(\ref{eq:dilatations_linear}) is again a surface term, in complete
analogy with~(\ref{eq:dilaton_Q_coupling}). It is nevertheless conceivable
that nonperturbative gravitational dynamics renders such terms physical,
thereby generating extra (manifestly non-QCD) pieces in the potential of
$\c$.\footnote{The schematic form of such contributions will be briefly
discussed in the next section.} Provided that any such gravitational
contributions, if present at all, are subdominant compared to the QCD-induced
potential, which we will be assuming, the following discussions go through
qualitatively unchanged.

We thus conclude that the exact form of non-QCD contributions, which are
expected to be present in one way or another, cannot be fixed by symmetry
alone and depends on the microscopic source of breaking. Regardless of that,
however, the low-energy effect is one and the same: it removes the degeneracy
among neighboring QCD ground states. For definiteness, and only as a proof of
principle, we parametrize it by adding to the
Lagrangian~(\ref{eq:Lagrangian_dilaton_exact}) a term
\be
\label{eq:dilaton_potential}
\delta V(\c) = \gamma \widetilde M^{4-p} \c^p \ ,
\ee
with $\gamma$ dimensionless, $\widetilde M$ a parameter with dimensions of
mass, and for the scaling dimension we require that $p\neq 4$.

\section{Nonlinearly realized scale symmetry}
\label{sec:nonlinear}

The theory in the broken phase\,\footnote{In cosmological backgrounds and for
sufficiently small explicit breakings, scale symmetry can be broken
dynamically~\cite{Ferreira:2018itt}, a mechanism coined ``inertial''
breaking.} is obtained by performing a Weyl rescaling of the metric
\be
g_{\m\n} \mapsto \f{M_{\rm Pl}^2}{\x_\c \c^2} g_{\m\n} \ ,
\ee
followed by the introduction of the canonical dilaton/axion $a$
\be
a = M_{\rm Pl}\, \x\log\left(\f{\c}{M}\right) \ ,
~~~\x = \sqrt{6+\f{1}{\x_\c}} \ ,
\ee
which yields
\be
\label{eq:dilaton_Einstein}
\mc L = \f{M_{\rm Pl}^2}{2}R  - \f 1 2 (\p_\m a)^2 
+ \f{a}{f_a} Q - \Lambda^4 e^{-q a/M_{\rm Pl}} 
- \f{\a M_{\rm Pl}^4}{\x_\c^2} \ , 
\ee
where 
\be
f_a = \f{\x}{\beta} M_{\rm Pl} \ ,
~~~\Lambda =  M_{\rm Pl}\left(\f{\widetilde M}{M}\right)^\f{4-p}{4}
\left(\f{\gamma}{\x_\c^2}\right)^\f 1 4 \ ,
~~~q = \f{4-p}{\x} \ .
\ee
Notice that the last term originating from the dilaton quartic piece of the
potential in~(\ref{eq:Lagrangian_dilaton_exact}), is a cosmological constant.

Several comments are in order here. First, it is the nonminimal coupling that
makes the broken-phase dilaton derivatively coupled, apart from the
exponential term,\footnote{The same functional form of the potential is
familiar from the original quintessence
literature~\cite{Wetterich:1987fk,Wetterich:1987fm,Ratra:1987rm}.}
in~(\ref{eq:dilaton_Einstein}). After the Weyl rescaling, the canonically
normalized field $a$ is logarithmic in $\c$, meaning that the theory
nonlinearly realizes scale invariance as (approximate) shift
symmetry.\footnote{The presence of a nonminimal coupling to gravity is
absolutely essential also in the Higgs-dilaton
model~\cite{Shaposhnikov:2008xb,Garcia-Bellido:2011kqb,Bezrukov:2012hx} and
its generalizations~\cite{Casas:2017wjh,Casas:2018fum}, so that the massless
in that case dilaton be only derivatively coupled to matter and thus evade
long-range (fifth) force constraints.} Note that one could equally well
consider~(\ref{eq:dilaton_Einstein}) as the starting point, impose the
approximate shift symmetry of $a$, and simply postulate that any explicit
breaking be exponentially small. The merit of the above derivation is that it
provides a concrete origin for this structure: the explicit breaking of scale
invariance translates into an exponential tilt for $a$, whose sign and
parametric size are controlled by the parent scaling dimension $p$ and the
nonminimal coupling $\x_\c$. 

Second, as emphasized in the previous section, the precise form of the non-QCD
contribution is not fixed by symmetry alone. For the benchmark choice $\d
V(\c)\propto \c^p$, relevant ($p<4$), irrelevant ($p>4$), nearly marginal
deformations ($p=4-\delta,~0<\delta\ll 1$), as well as integration-constant
($p=0$) sources of breaking, all lead to an exponential breaking of the shift
symmetry of $a$. The type of the deformation~(\ref{eq:dilaton_potential})
determines the sign and partly sets the parametric size of the slope through
the $(4-p)$ factor. As anticipated, for $p=4$ one sees that $q=0$ and the
potential collapses into a cosmological constant, independent of $a$; hence it
does not lift the degeneracy of the QCD vacua and cannot provide the required
bias. In contrast, for any $p\neq 4$, the form of the tilt is universal. Its
overall magnitude, however, can be further adjusted by the nonminimal coupling
$\x_\c$. In particular, even for a strongly relevant deformation ($p\to 0$)
the exponential can be made shallow for sufficiently small $\x_\c$, since then
$\x\simeq 1/\sqrt{\x_\c}\gg 1$ and $q\propto 1/\x \ll 1$.

This point is important because different considerations can motivate
different parametric expectations for $\x_\c$. On the one hand, naive
dimensional analysis suggests that a nonminimal coupling can be as large as
$\mc O(16\pi^2)$, following from requiring that the UV cutoff of the
gravitational sector $\sim 4\pi M_{\rm Pl}/\sqrt{\x_\c}$ be comparable to the
Planck scale~\cite{Csaki:2014bua}.\footnote{Larger values for a nonminimal
coupling are actually more interesting phenomenologically, as they allow the
Higgs field to inflate the universe~\cite{Bezrukov:2007ep}.}~On the other
hand, in concrete realizations such as the Higgs-dilaton
model~\cite{Shaposhnikov:2008xb,Garcia-Bellido:2011kqb}, where the tilt is due
to the unimodular integration constant (corresponding to $p= 0$), the
inflationary dynamics and CMB constraints require that $\x_\c\ll 1$. Put
differently, for fixed $p$ the tilt becomes shallower as $\x_\c$ decreases,
while for large $\x_\c$ one approaches the limiting value $\x\to\sqrt{6}$ and
$q\sim \mc O(1)$, for all reasonable choices of scaling dimensions. For a QCD
axion with $f_a\ll M_{\rm Pl}$, which in the present context corresponds to
$\b \gg |4-p|$, the combination that controls the variation of the tilt across
adjacent QCD branches 
\be
\label{eq:epsilon}
\epsilon = q\frac{f_a}{M_{\rm Pl}} = \f{4-p}{\b} \ ,
\ee
is typically very small, and so is the breaking of the shift symmetry. 

Third, for $q<0$ (corresponding to irrelevant parent operator) the exponential
runs in the opposite direction. Since the subsequent defect history is simply
obtained by $a\to -a$, throughout we will take $q>0$ for definiteness.

Before moving on, our last comment concerns the form of the contributions to
the potential due to the possible couplings~(\ref{eq:Jordan_topological}) of
the dilaton to the gravitational topological densities. A heuristic way to
think about them is the following. Once in the Einstein frame and working with
the canonical field $a$, the topological pieces of the
Lagrangian~(\ref{eq:Jordan_topological}) become\,\footnote{The Weyl rescaling
of $E_4$ generates higher-derivative terms involving $a$ and curvatures. These
are not relevant for our considerations here, as they are not expected to
alter the energy splitting of vacua controlling the collapse of domain walls.}
\be
\mc L _{\rm grav} \supset \f{a}{f_E} E_4 + \f{a}{f_P} P_4 \ ,
\ee
meaning that the couplings of $a$ to the densities are linear in the canonical
field; $f_E,f_P$ are dimensionful parameters. Despite their similarities, the
Euclidean continuation suggests a natural distinction between them.

Indeed, the Euler density contains two Levi-Civita symbols, whereas the
Pontryagin density contains only one. Upon analytic continuation, this implies
that the $E_4$ term contributes as a real deformation of the Euclidean action,
while the $P_4$ term contributes as a purely imaginary phase, much like
ordinary QCD. One therefore expects that saddles with nonzero Euler number
generate contributions weighted schematically by $e^{-a/f_E}$ or sums thereof,
whereas saddles with nonzero Pontryagin number contribute as
$e^{i(a/f_P+\theta_{\rm grav})}$ with $\theta_{\rm grav}$ a possible
gravitational CP-odd phase. This should produce two qualitatively different
structures in the effective potential
\be
\label{eq:gravitational_NP_lifting}
V_{E_4}(a)\sim \sum_i \Lambda_{E,i}^4\,e^{-a/f_{E,i}} \ ,
~~~V_{P_4}(a)\sim \sum_i \Lambda_{P,i}^4
\cos\l(\f{a}{f_{P,i}}+\theta_{\rm grav,\,i}\r) \ ,
\ee
or suitable generalizations and combinations thereof. In this sense, the Euler
sector is associated with extra nonperiodic exponential contributions, while
the Pontryagin sector may induce oscillatory pieces in the potential. 

For sufficiently large $f_{P,i}$, the second term
in~(\ref{eq:gravitational_NP_lifting}) may be expanded locally around a
minimum. For a single contribution, this gives 
\be
V_{P_4}(a) \approx {\rm const} + \f 1 2 m_{P}^2 a^2+\ldots \ ,
~~~m_P^2\sim \f{\Lambda_P^4}{f_P^2}\ .
\ee
Therefore, for $a/f_{P,i}\ll 1$, the Pontryagin sector can mimic the quadratic
non-compact tilt considered in~\cite{Karananas:2025uhy}. By contrast, the
Euler-type contributions $V_{E_4}(a)$ lead to nonperiodic exponential terms of
the form used in~(\ref{eq:dilaton_potential}).

Although very appealing, as this would provide a fully gravitational tilting
of the potential, we stress that how, and most importantly whether, this even
materializes is far less understood than the corresponding QCD dynamics, so we
will not be relying on~(\ref{eq:gravitational_NP_lifting}). Nevertheless, this
discussion provides argument for why gravitational topological sectors may
generate precisely the sort of non-QCD contributions relevant for the present
framework. 

The Lagrangian~(\ref{eq:dilaton_Einstein})---supplemented with an
(appropriate) inflationary sector---contains all the ingredients needed for
the noncompact QCD axion dynamics to be operative. In particular, $a$ is
coupled to the QCD topological density, while its exponential potential lifts
the degeneracy among adjacent periodic vacua that appear once QCD confines and
thereby destabilizes the domain-wall network. Moreover, the tilt induces CP
violation.

\section{Phenomenological consequences}
\label{sec:pheno}

\begin{figure}
\centering
\includegraphics[scale=.4]{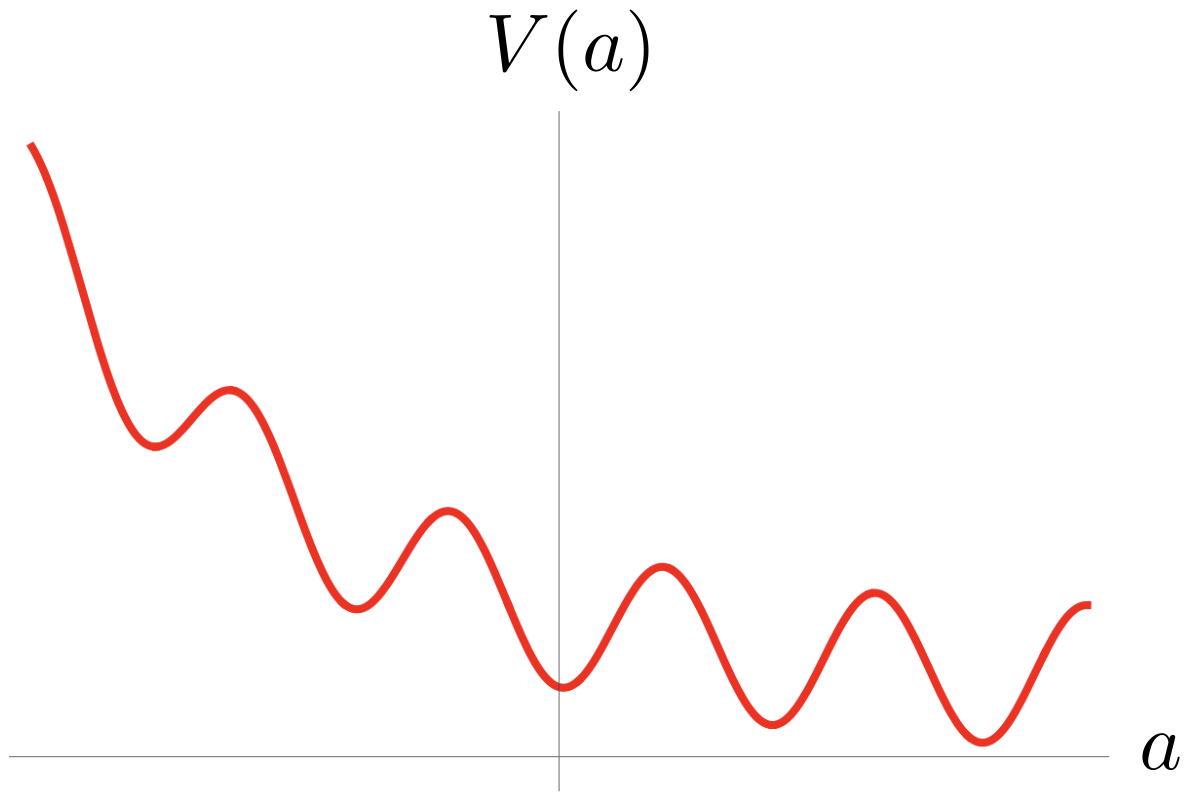}
\caption{Axion potential in the presence of the (grossly exaggerated for
illustration purposes) exponential tilt. The usual periodic QCD contribution
would give degenerate minima at $a=2\pi n f_a$. The monotonic non-QCD
contribution lifts this degeneracy, and shifts each minimum by the small
branch-dependent amount~(\ref{eq:theta_phys}) that corresponds to the residual
strong-CP violation in that vacuum.}
\label{fig:potential}
\end{figure}

The picture painted so far maps directly onto the non-compact axion cosmology
of~\cite{Karananas:2025uhy}. We briefly summarize the main consequences,
emphasizing the aspects that are specific to the exponential bias and
contrasting them with the previously studied quadratic tilt.

The full axion potential consists of the usual QCD periodic piece and the
exponential bias. The resulting ``tilted washboard'' structure is shown
schematically in Fig.~\ref{fig:potential}. Expanding around the $n$-th QCD
branch, whose minimum in the absence of the tilt is at $a/f_a \simeq 2\pi
n+\bar\theta$, one finds
\be
\label{eq:full_potential}
V(a)\approx \f{f_a^2 M_a^2}{2}
\left(\f{a}{f_a}-2\pi n-\bar\theta\right)^2
+\Lambda^4 e^{-\epsilon a/f_a} \ ,
\ee
where~\cite{ParticleDataGroup:2024cfk}
\be
\label{eq:QCD_mass}
M_a \simeq 5.7\times10^{-15}~{\rm GeV}
\left(\f{10^{12}~{\rm GeV}}{f_a}\right)
\ee
is the zero-temperature QCD-induced axion mass, and $\epsilon$ was introduced
in~(\ref{eq:epsilon}).

Minimizing~(\ref{eq:full_potential}) within the branch labeled by $n$ gives a
residual strong CP phase
\be
\label{eq:theta_phys}
\theta_{\rm phys}\simeq
\epsilon\left(\f{\Lambda^2}{f_a M_a}\right)^2
e^{-\epsilon(2\pi n+\bar\theta)} \ ,
\ee
which must satisfy the neutron electric-dipole-moment
bound~\cite{Abel:2020pzs}
\be
\label{eq:theta_bound}
\theta_{\rm phys}\lesssim 10^{-10} \ .
\ee
Thus, the very same non-QCD tilt needed to lift the vacuum degeneracy
necessarily induces a nonzero residual strong phase. Equivalently, the energy
difference between neighboring vacua can be written as
\be
\label{eq:pressure_theta}
\Delta V \simeq 2\pi f_a^2 M_a^2\,\theta_{\rm phys} \ .
\ee
This relation is particularly useful, as it ties the cosmological wall bias
directly to a low-energy observable.

Before turning to the wall dynamics, it is worth extracting from
(\ref{eq:theta_phys}) a useful estimate for the overall scale of the non-QCD
contribution. Since $\epsilon\ll1$, the exponential factor varies only mildly
over the populated range of branches, so $\theta_{\rm phys}$ fixes primarily
the combination $\epsilon(\Lambda^2/f_a M_a)^2$. One then finds that 
\be
\label{eq:Lambda_estimate}
\Lambda \sim (f_a M_a)^{1/2}
\left(\f{\theta_{\rm phys}}{\epsilon}\right)^{1/4}
\sim \mc O(10)~{\rm MeV} \ ,
\ee
for the phenomenologically relevant region of parameter space, see below. 

There is, however, a mild constraint on the value of the $a$ field. Strictly
speaking, the relevant quantity is not $\Lambda$ alone, but the size of the
exponential bias over the set of QCD branches populated during inflation. In
particular, for the branches relevant for the final wall collapse, with field
value denoted by $a_c$, one should require
\be
2.3\times10^{-11} \l(\f{f_a}{1.4\times10^{11}~{\rm GeV}}\r)
\lesssim
\epsilon\left(\f{\Lambda^2}{f_a M_a}\right)^2 e^{-\epsilon a_c/f_a}
\lesssim
10^{-10} \ ,
\ee
where the upper bound is the electric-dipole-moment constraint and the lower
bound is the value required for sufficiently early wall annihilation,
estimated below in~(\ref{eq:theta_window}). Thus, for $q>0$, the homogeneous
value of the field cannot be arbitrarily large and positive: otherwise the
exponential bias becomes too small and the walls are long-lived. Conversely, a
too negative value of $a_c$ enhances the bias and can violate the EDM bound or
spoil the assumption that the QCD potential controls the local structure of
the minima.

The cosmological requirements of the non-compact mechanism can now be
summarized rather economically. First, inflation must populate many QCD
branches. This requires the inflationary fluctuations to exceed the spacing
between neighboring QCD minima~\cite{Karananas:2025uhy}
\be
\label{eq:fluctuations_spread}
H_I \gtrsim 2\sqrt{2\pi}\,f_a \ ,
\ee
so that different Hubble patches relax to different vacua once QCD turns on.

A further consistency check is that the exponential tilt should not already
induce an appreciable classical drift of the field during inflation, since the
stochastic spread must dominate the evolution. Quantitatively, the quantum
fluctuation per inflationary efolding, $\sim H_I/2\pi$, should exceed the
corresponding classical displacement,
\be
\label{eq:fluctuations_vs_slow-roll}
\frac{H_I}{2\pi}\ \gtrsim\ \frac{|V'(a_I)|}{3H_I^2}
\ \simeq\ \f{\epsilon\Lambda^4}{3f_a H_I^2} e^{-\epsilon a_I/f_a} \ ,
\ee
where $a_I$ denotes the value of the field during inflation. Using
Eqs.~(\ref{eq:QCD_mass},\ref{eq:theta_phys}), this may be rewritten as
\be
\label{eq:fluctuations_vs_slow-roll_numeric}
H_I \gtrsim 4.1\times 10^{-6}\,\theta_{\rm phys}^{1/3}
\left(\f{10^{12}~{\rm GeV}}{f_a}\right)^{1/3}
e^{-\frac{\epsilon}{3}\left(\frac{a_I}{f_a}-2\pi n-\bar\theta\right)}
~{\rm GeV} \ .
\ee
In practice, once $\theta_{\rm phys}$ is required to
satisfy~(\ref{eq:theta_bound}), this condition is automatically met for any
reasonable inflationary history. It therefore does not impose an additional
meaningful restriction on the parameter space.

Second, the effective theory must remain under control during inflation. 
Tree-level unitarity of processes mediated by the axionic coupling to gluons
implies a cutoff~\cite{Karananas:2025uhy} $\Lambda_I\sim
8\pi^{3/2}f_a/\alpha_s $, and therefore
\be
\label{eq:HI_window}
\f{8\pi^{3/2}}{\alpha_s}f_a \gtrsim H_I \gtrsim 2\sqrt{2\pi}\,f_a \ .
\ee
Together with the observational upper bound on the inflationary Hubble
scale~\cite{Planck:2018jri}, this singles out high-scale inflation, \eg
Higgs~\cite{Bezrukov:2007ep}, Starobinsky~\cite{Starobinsky:1980te}, and
$\alpha$-attractors~\cite{Kallosh:2013lkr,Kallosh:2013yoa,Galante:2014ifa}.
Roughly, 
\be
H_I\sim \mc O(10^{12-13})~{\rm GeV} \ ,
\ee
for the range of $f_a$ relevant to the mechanism. In this regime the axion
typically explores
\be
N\sim \f{H_I}{f_a}\sim \mc O(10\text{--}10^3) \ ,
\ee
distinct QCD branches, so that the largest populated branch index is
parametrically of the same order, $n_{\rm max}\sim N$; in the
above estimate we dropped factors of $2\pi$ as well as the mild (square root)
dependence of the axion perturbation on the number of efoldings between the
relevant epochs.

Once QCD turns on, a complicated network of domain walls forms between regions
that relax to neighboring branches of the potential. The bias
(\ref{eq:pressure_theta}) destabilizes this network and causes it to
annihilate~\cite{Sikivie:1982qv,Sikivie:2006ni}. For the exponential tilt, one
finds
\be
\label{eq:pressure_explicit}
\Delta V \simeq
2\pi\epsilon\Lambda^4 e^{-\epsilon(2\pi n+\bar\theta)} \ ,
\ee
so that the pressure decreases mildly with $n$. At the same time, for
$\epsilon>0$ the vacuum energy itself also decreases with $n$, so larger-$n$
branches are deeper but less strongly biased relative to their neighbors. As a
result, the walls collapse almost together, with the last ones to disappear
being those adjacent to the deepest populated ground state. This should be
contrasted with the quadratic tilt~\cite{Karananas:2025uhy}, for which the
branch energy grows quadratically with $n$ while the bias between neighboring
branches increases with $n$. There too the last surviving walls border the
deepest populated vacuum, but the relation between branch depth and branch
index is reversed. 

Estimating the annihilation time as
\be
t_\ast \sim \f{\sigma_{\rm DW}}{\Delta V} \ ,
\ee
with $\sigma_{\rm DW}\simeq 9 M_a
f_a^2$~\cite{Sikivie:2006ni,GrillidiCortona:2015jxo}, and requiring that the
walls disappear before Big Bang Nucleosynthesis, one obtains a lower bound on
the residual strong phase,
\be
\label{eq:theta_window}
\theta_{\rm phys} \gtrsim 2.3\times10^{-11}
\l(\f{f_a}{1.4\times10^{11}~{\rm GeV}}\r) \ .
\ee
Combined with the upper bound~(\ref{eq:theta_bound}), this leads to a narrow
predictive window for strong CP violation. If the axion accounts for all of
the dark matter, corresponding roughly to the benchmark $f_a\simeq
1.4\times10^{11}~{\rm GeV}$, the allowed window is only a factor of few wide.
This should also be contrasted with the quadratic tilt
of~\cite{Karananas:2025uhy}, for which the corresponding lower bound on
$\theta_{\rm phys}$ is roughly forty times smaller than the current upper
bound. In this sense, the exponential bias is more predictive. It should be
stressed that the above estimate neglects axions emitted during the collapse
of the wall network. These are expected to provide an additional contribution
to the dark matter relic abundance. Depending on the efficiency of this
emission, lower values of $f_a$ may be needed in order to not overclose the
Universe.

Note that the use of the astrophysical
constraint~\cite{Raffelt:1990yz,Caputo:2024oqc,Fiorillo:2025gnd,Candon:2025sdm}
$f_a \gtrsim \mc O (10^9)$ GeV---which is strongly model-dependent on how the
axion couples to fermions---weakens the bound on $\theta_{\rm phys}$ by two
orders of magnitude, $\theta_{\rm phys} \gtrsim 2\times10^{-13}$, but still
leaves it accessible in the future experiments on the electric dipole
moment~\cite{Anastassopoulos:2015ura,Grasdijk:2020ihi,pEDM:2025nlu}. 

The collapse of the wall network also sources a stochastic background of
gravitational waves. Following the estimate of~\cite{Karananas:2025uhy}, and
assuming roughly one wall per Hubble patch~\cite{Sikivie:1982qv}, the
characteristic frequency at production is set by the annihilation time,
\be
\omega_\ast \sim t_\ast^{-1} \ ,
\ee
while the fractional energy density at emission scales as
\be
\Omega_{{\rm GW},\ast}\simeq
2\,\theta_{\rm phys}^{-2}\left(\f{f_a}{M_{\rm Pl}}\right)^4 \ .
\ee

Redshifted to the present epoch, and for the benchmark parameters relevant to
the dark-matter motivated range of $f_a$, the signal lies in the nanohertz
band 
\be
\omega_0\sim {\rm nHz} \ ,
\ee
and has amplitude of order
\be
\Omega_{{\rm GW},0}\sim \mc O (10^{-11}{\rm-}10^{-12}) \ .
\ee
Such a background could fall inside the frequency window targeted by future
pulsar timing arrays~\cite{Babak:2024yhu}.

\section{Conclusion}
\label{sec:conclusion}

We have shown that the (pseudo-)Nambu--Goldstone mode of spontaneously broken
approximate scale invariance provides a remarkably economical realization of
the non-compact QCD axion. The key structural ingredient is the nonminimal
coupling to gravity. In the broken phase it renders the canonically normalized
field derivatively coupled, so that the theory nonlinearly realizes scale
invariance as an approximate shift symmetry. Small explicit breakings of scale
symmetry then translate universally into exponentially small breakings of the
shift symmetry and automatically generate the tiny non-QCD tilt required for a
viable defect history. The non-compactness of the field enables it to sample
many QCD branches after confinement, isocurvature is efficiently washed out,
and the ensuing domain-wall network---unaccompanied by cosmic strings---is
biased to annihilate before BBN. The same bias that destabilizes the walls
shifts the vacuum away from the CP-conserving point, yielding a narrow and
predictive range of residual strong-CP violation, correlated with a
nanohertz-peaked stochastic gravitational-wave signal from wall annihilation.
Future electric-dipole-moment searches and pulsar-timing arrays provide a
distinctive probe of the framework. 

A concrete realization is furnished by the Higgs--dilaton
model~\cite{Shaposhnikov:2008xb,Garcia-Bellido:2011kqb} with unimodular
gravitational dynamics, where the required runaway potential arises in the
gravitational sector as an integration constant and no additional explicit
scale-breaking operators are radiatively generated when the theory is
renormalized in a scale-invariant
manner~\cite{Englert:1976ep,Shaposhnikov:2008xi}. In that case, the dilaton
inherits precisely the type of exponential tilt needed for the non-compact
axion cosmology, while remaining protected by the approximate shift symmetry
originating from exact scale invariance. At the same time, inflation is due to
the Higgs field, thereby providing a well-motivated and in excellent agreement
with observations~\cite{Planck:2018jri,BICEP:2021xfz} cosmological dynamics.  

A final remark concerns the broader role of the exponential potential. In the
present work we focused on the original non-compact axion cosmology, where the
exponential acts as a small bias that lifts the degeneracy among neighboring
QCD vacua. The same functional form, however, also appears in the ``scaling
QCD axion'' scenario~\cite{preparation}, in which the runaway potential
severely affects the pre-QCD evolution of the field. Such a parameter regime
does not seem generic in the present setup. Here, the exponential is assumed
to be subdominant to the QCD potential at the epoch relevant for vacuum
selection, so its natural role is to provide a small bias rather than to
control the earlier homogeneous evolution. Thus, despite the common functional
form, the resulting dynamics differs crucially.

\section*{Acknowledgements} 
It is a great pleasure to thank Sebastian Zell for comments on the manuscript.

\bibliographystyle{utphys}
\bibliography{Refs}

\end{document}